\documentclass[doi-false, twocolumn,aps,amsmath,nofootinbib,pre,floatfix,superscriptaddress]{revtex4-2}
\usepackage[utf8]{inputenc}
\usepackage{lineno}
\modulolinenumbers[5]
\usepackage{float}
\usepackage{comment}
\usepackage{graphicx}
\usepackage[export]{adjustbox}
\usepackage{wrapfig}
\usepackage{lipsum}  
\usepackage{amsmath}
\usepackage{amssymb}
\usepackage{color}
\usepackage{mdframed}
\usepackage[normalem]{ulem}

\usepackage{hyperref}
\hypersetup{
    colorlinks=true,
    linkcolor=blue,
    filecolor=magenta,      
    urlcolor=cyan,
}

\usepackage{xcolor}
\definecolor{mendes_green}{rgb}{0,0.5,0.12}
 
\begin{document}
\title{Cooperation in regular lattices}

\author{Lucas S. Flores}
\affiliation{Instituto de Física, Universidade Federal do Rio Grande do Sul, CP 15051, CEP 91501-970 Porto Alegre - RS, Brazil}
\author{Marco A. Amaral}
\affiliation{Instituto de Humanidades, Artes e Ciências, Universidade Federal do Sul da Bahia, CEP 45638-000, Teixeira de Freitas - BA, Brazil.}
\author{Mendeli H. Vainstein}
\affiliation{Instituto de Física, Universidade Federal do Rio Grande do Sul, CP 15051, CEP 91501-970 Porto Alegre - RS, Brazil}
\email[]{vainstein@if.ufrgs.br}
\author{Heitor C. M. Fernandes}
\affiliation{Instituto de Física, Universidade Federal do Rio Grande do Sul, CP 15051, CEP 91501-970 Porto Alegre - RS, Brazil}
\email[]{heitor@if.ufrgs.br}

\date{\today}

\begin{abstract}

In the context of Evolutionary Game Theory, one of the most noteworthy mechanisms to support cooperation is spatial reciprocity, usually accomplished by distributing players in a spatial structure allowing cooperators to cluster together and avoid exploitation. This raises an important question: how is the survival of cooperation affected by different topologies? 
Here, to address this question, we explore the Focal Public Goods (FPGG) and classic Public Goods Games (PGG), and the Prisoner's Dilemma (PD) on several regular lattices:  honeycomb, square (with von Neumann and Moore neighborhoods), kagome, triangular, cubic, and 4D hypercubic lattices using both analytical methods and agent-based Monte Carlo simulations. 
We found that for both Public Goods Games, a consistent trend appears on all two-dimensional lattices: as the number of first neighbors increases, cooperation is enhanced. However, this is only visible by analysing the results in terms of the payoff's synergistic factor normalized by the number of connections. 
Besides this, clustered topologies, i.e., those that allow two connected players to share neighbors, are the most beneficial to cooperation for the FPGG. 
The same is not always true for the classic PGG, where having shared neighbors between connected players may or may not benefit cooperation. We also provide a reinterpretation of the classic PGG as a focal game by representing the lattice structure of this category of games as a single interaction game with longer-ranged, weighted neighborhoods, an approach valid for any regular lattice topology.
Finally, we show that depending on the payoff parametrization of the PD, there can be an equivalency between the PD and the FPGG;  when the mapping between the two games is imperfect, the definition of an effective synergy parameter can still be useful to show their similarities.

\end{abstract}
\maketitle

\graphicspath{{figs/}}


\section{Introduction}

The emergence and maintenance of cooperative behavior in a competitive environment are one of the most long-standing issues in biological and social sciences~\cite{Pennisi2005, Nowak2011a, galam_ijmpc08, Capraro2018, rand_tcs13}. After all, why would self-interested agents pay a cost to provide benefits to others?
And yet, cooperation permeates nature, being much more common than could be anticipated based on a naive application of the Darwinian premise that only the fittest individuals survive. Humans, social insects (such as bees and ants), flocks of birds, and even members of different species can mutually cooperate and share benefits~\cite{Perc2017, wilson_71, Nowak2011a}. 
It is no surprise, therefore, that much research has been dedicated to discovering possible mechanisms that support cooperation~\cite{rand_tcs13, Perc2017,buchan_pnas09, galam_ijmpc08, Capraro2018, Szabo2007, perc_bs10}. 
Evolutionary Game Theory (EGT) is one of the most useful mathematical frameworks for dealing with such questions~\cite{Smith82, Axelrod1984, Weibull1995, nowak_s06}. Its applications range from economics~\cite{jiang_ll_pone13} to epidemiology~\cite{Amaral2020d, Jentsch2021, Kabir2021}, rumor spreading~\cite{Amaral2018b}, the evolution of moral behavior~\cite{Kumar2020}  and even quantum mechanics~\cite{Vijayakrishnan2020}.  
By resorting to games such as the famous Prisoner's Dilemma and the Public Goods Game, the EGT framework can mathematically model situations where the best outcome for the whole population occurs when everybody cooperates, while at the same time the best choice for the individual is to betray their peers, obtaining all the benefits without having to pay a cost. This encapsulates the dilemma between individual and collective gains and can be modeled in an evolutionary setting where more efficient strategies leave more offspring.
Classic examples reviewed in~\cite{nowak_s06} include kin selection~\cite{maynard_n64}, direct and indirect reciprocity~\cite{trivers_qrb71, axelrod_s81}, network reciprocity~\cite{Nowak1992a, wardil_epl09, wardil_pre10, wardil_jpa11, NagChowdhury2020, Vukov2012, Wu2018a}, as well as group selection~\cite{wilson_ds_an77, Javarone2017}. Diffusion and mobility have also been studied prominently~\cite{Vainstein2001a, vainstein_jtb07, vainstein_pa14}, as were various coevolutionary models~\cite{perc_bs10} involving network topology, noise~\cite{Amaral2020a,Amaral2018,Amaral2020}, heterogeneity~\cite{Zhao2020, Sendina-Nadal2020, santos_jtb12, Sparrow1999, fort2008minimalevol, Szolnoki2018a, Amaral2016, Amaral2015, Tanimoto2017} and aspiration~\cite{Amaral2016b, pacheco_prl06, pacheco_jtb06, wu_zx_pre06, pestelacci_lncs08, tanimoto_pa13, biely_pd07, cardillo_njp10, dai_ql_cpl10, hetzer_pone13}. 
Among the mechanisms that promote cooperation, perhaps one of the most well-studied is the reciprocity that emerges from the spatial distribution of players. In a novel approach at the time, Nowak and May~\cite{Nowak1992a} modeled spatially distributed populations where players copy the fittest strategy nearby. By doing so, cooperators can form compact clusters supporting themselves in a sea of defectors.
Nevertheless,  structured populations are not always favorable for the maintenance of cooperation~\cite{ Hauert2004,gracia-lazaro_pnas12}, and the specific type of spatial topology can drastically alter the population dynamics~\cite{santos_prsb06}. For example, while the snowdrift game presents trivial dynamics in square lattices, the triangular lattice creates topological frustration akin to glassy systems in condensed matter, leading to long-range strategy ordering, a phenomenon absent in most lattices~\cite{Amaral2017}.

Aiming to better understand the effects of diverse topologies on cooperation, here we comprehensively explore both the focal Public Goods (FPGG) and  classic Public Goods (PGG) Games, and the Prisoner's Dilemma (PD) in some of the most common two-dimensional regular topologies: the square, triangular, kagome and honeycomb lattices. For the sake of completeness, we also compare the results with the cubic and $4D$ hypercubic lattices. 
 While the square lattice has been consistently explored, the lack of systematic study of different regular lattices in the literature for a given game and set of parameters is evident. This work proposes to fill this gap and put all systems under consideration under well-established analyzes.  We also deem important to note that complex networks such as the small-world~\cite{watts_dj_n98} and scale-free networks~\cite{Barabasi1999} have been thoroughly studied in the framework of evolutionary game theory. However, the scope of the current paper is to focus mainly on regular lattices and to compare the effects of those in some of the most usual games.

We hope that this work will provide a robust basis for comparing the effect of different topologies in diverse games, creating a unified framework to compare different lattices and their effect on these games. We also include an analytical reinterpretation of the PGG, showing that we can represent a lattice structure of such category of games as a single interaction game with longer-ranged, weighted interactions, an approach valid for any regular lattice topology.

\section{Models}
\label{model}

\begin{figure*}[htb]
    \centering
    \includegraphics[width=\linewidth]{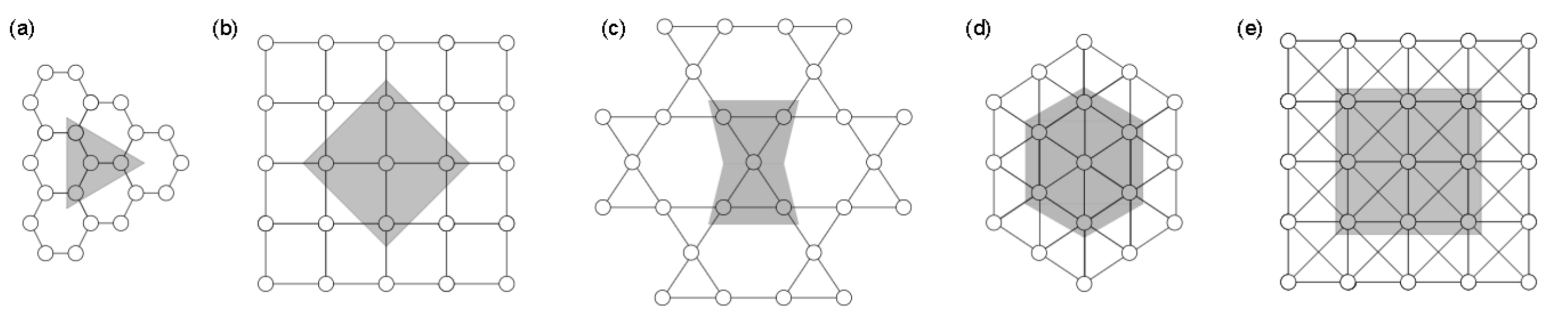}
    \caption{Illustrations of segments of (a) honeycomb, $G=4$, (b) square with von Neumann neighborhood, $G=5$, (c) kagome, $G=5$, (d) triangular, $G=7$, and (e) square with Moore neighborhood, $G=9$, lattices. The vertices represent the players and the edges are the connections between them. We highlight all first neighbors that form the group in which the central player is the focal site for each topology, i.e., all vertices directly linked to the focal site. Simulations on cubic and $4D$ hypercubic lattices were also performed.}
    \label{topo}
\end{figure*}

We will study three different models: the Prisoner's Dilemma (PD), the Focal Public Goods Game (FPGG), and the classic Public Goods Game (PGG).
For the FPGG, the focal player and its nearest neighbors form a group of size $G$. Each player can choose between cooperating ($C$) by contributing with an investment $c$ to a public pool, or defecting  ($D$) without contributing. The sum of all contributions is multiplied by a synergistic factor $r>1$ and then equally distributed among all players in the group, including  defectors. We stress that the group size $G$, the number of nearest neighbors, and the total population size $N$, will depend on the lattice topology (see Fig.~\ref{topo}. 
As usual for the FPGG, the agent's payoff consists only of the earnings from the group in which it is the central site, and is given by
\begin{eqnarray}
\Pi_D&=&\frac{rc}{G} N_C, \label{payD}\\
\Pi_C&=&\frac{rc}{G} (1+N_C) -c, \label{payC}
\end{eqnarray}
where $\Pi_C$ ($\Pi_D$) is a cooperator's (defector's) payoff, and $N_C$ is the number of cooperative players around the player in question.  
For the classic PGG, on the other hand, a player's total payoff is equal to the sum of earnings from all $G$ groups to which the player belongs. 

The population dynamics is implemented via a Monte Carlo simulation for all games. Each Monte Carlo trial movement is comprised of the following steps: first, a random player $i$ is selected and its payoff is calculated. Then, this process is repeated with a randomly selected neighbor $j$. Next, player $i$ tries to imitate player $j$'s strategy with probability:
\begin{equation} \label{eq.transition}
     W_{i\rightarrow j}= \frac{1}{1+e^{-(\Pi_j-\Pi_i)/K}} \,\,,   
\end{equation}
where $ K $ is the noise associated with irrationality during interactions~\cite{Szabo2007, szolnoki_pre09c}.
The previous process is repeated $N$ times, characterizing one Monte Carlo step (MCS). This is done to guarantee that, on average, each player will have a chance to change their strategy at each MCS. We present results for $t_{max} = 10 ^ 5\, \text{MCS} $.
We use periodic boundary conditions for all regular lattices presented. Initially, we randomly distribute the strategies with half of the players being defectors $( D )$ and the other half, cooperators $(C)$.
In the simulations, we used $K=0.1$ and averaged over $100$ independent samples to generate the results, unless otherwise stated. 

It is important to notice the role of the investment cost $c$ in eqs.~\ref{payD} and \ref{payC}.
Since all terms in both payoffs are proportional to $c$, a change in its value can also be viewed as a rescaling of the noise $K$ ($K^\prime = K/c$) in eq.~\ref{eq.transition}.
Therefore, a scenario with a high contribution value is equivalent to a low noise case and vice-versa. 
The usual approach of setting the agent's contribution cost to unity ($c=1$), which we have followed, is equivalent to using the rescaled noise $K'$ making the dynamics independent of the particular choice of $c$. This observation will be crucial for the comparison of the PD and the FPGG in section~\ref{pd_sec}.

While both PGG and PD games can encapsulate the social dilemma of collaborating in a selfish environment, they have small differences, the central one being that the PD is a $1\times 1$ interaction game, i.e., the games are played between only two players at a time. On the other hand, the PGG is more suitable for model situations where each individual will play at the same time against a group of different agents. However, in a spatial setting, the distinction between them becomes subtler, and for some cases, one can even be mapped onto the other~\cite{Hauert2003}. 

In the PD, two players can either cooperate $(C)$ or defect $(D)$: mutual cooperation yields a payoff $R$ (reward), and mutual defection yields $P$ (punishment) for both players. If players have different strategies, the defector receives $T$ (temptation) while the cooperator receives a small payoff $S$ (sucker). For the game to classify as a PD~\cite{Nowak1992a, Szabo2007, perc_bs10}, the payoffs should obey $2R>T+S$ and the hierarchy $T>R>P>S$. 
In classical game theory, defection is the Nash equilibrium and, therefore, the rational choice.
 The dynamics of the PD are the same as that for the FPGG: two randomly chosen neighboring sites play the game with all their first neighbors and themselves to obtain a payoff composed of the sum of the results from all games. The transition probabilities between strategies also follow eq.~\ref{eq.transition}.
A commonly used parametrization is  $T = \beta$, $R =  \beta - \gamma$, $P = 0$ and $S = -\gamma$ for the PD, which we chose so that we can compare the PD and the FPGG games, S done in detail in section~\ref{pd_sec}.


\subsection*{Topologies}
\label{sec.topologies}

We present in Fig.~\ref{topo} an illustrative diagram of all the two-dimensional structured neighborhoods used in our simulations (the $3D$ cubic and $4D$ hypercubic lattices are not shown), where each player is located at a vertex of a graph and the edges denote a given player's connections.
Each lattice under consideration has a specific primitive cell, as pointed out in Ref.~\cite{newmanb99}. 
We denote by $L$ the linear size of the lattice and the total population size by $N$. We will use the following lattices: honeycomb ($G=4$, $N = 2 \times L^2=2 \times 100^2$), kagome ($G=5$, $N=3 \times L^2=3 \times 100^2$),  triangular ($G=7$, $N = L^2=100^2$), cubic with von Neumann neighborhood ($G=7$, $N = L^3=20^3$), $4D$ hypercubic with von Neumann neighborhood ($G=9$, $N=L^4=10^4$), and square lattice with both the von Neumann ($G=5$, $N=L^2=100^2$) and the Moore ($G=9$, $N=L^2=100^2$) neighborhoods.
While the square lattice is a defined topology, the von Neumann and the Moore neighborhoods are different definitions of interacting neighbors in a given topology. In the current work, we use both types of neighborhoods when studying the square lattice. From now on, when we refer to the von Neumann and Moore neighborhoods, we are considering only the case of the square lattice.

It is important to emphasize that neighboring sites share a different number of neighbors (closed loops inside the group) depending on the lattice topology considered, therefore possessing a varying clustering coefficient~\cite{watts_dj_n98}. Some topologies, such as the square and kagome lattices, have the same number of neighbors ($G=5$) but a different number of shared neighbors. 
The latter has closed loops inside a given group while the former does not, see Fig ~\ref{topo}.
Specifically, there are no shared neighbors between two connected players for the honeycomb, square,   cubic, and $4D$ hypercubic lattices with von Neumann neighborhoods. On the kagome lattice, there is one; on the triangular lattice there are two; and on the square lattice with the Moore neighborhood, there are either two or four shared neighbors, depending on the connection direction.

\section{Results}
\label{Cooperação}

\subsection{Focal Public Goods Game}
\label{sec.FPGG}

\subsubsection*{Random neighborhood}
\label{sec.random}

In the well-mixed approximation, each player interacts with all other players in the population~\cite{Szabo2007}.
By doing so, all sites become equivalent and spatial effects are suppressed, once every player has the whole population as its neighborhood. This results in the extinction of cooperation for all parameter regions in all studied systems.

A different approach adopted here that mimics the well-mixed scenario is to maintain each player's group size equal to $G$ and to, at each time step, randomly select all other $G-1$ neighbors from the entire population. Note that for this case, the interacting sites may or may not participate in each other's group. 
This constant randomization diminishes the effects of spatial reciprocity and mimics a fully connected topology while preserving the dependence on $G$. 
As expected, when spatial reciprocity is suppressed by the constant mixing of interactions, cooperators are not able to establish compact clusters to support themselves. In this situation, there can only be full defection or full cooperation in the population, depending on $r$. Without spatial organization, both strategies are unable to co-exist.
In this setting, cooperation only survives for $r>G$ for both the FPGG and the classic PGG (data not shown).
For these two games, above the threshold value, a situation known as weak altruism~\cite{Wardil2017, Wilson1990} settles in: a cooperator always receives a positive return for their initial investment even when exploited by all other players and, therefore, cooperation thrives.

For the FPGG we can understand this result by analysing  eqs.~\ref{payD} and \ref{payC}. A cooperator has a higher payoff than a defector when $\Pi_C > \Pi_D$, giving the transition values
\begin{equation}
\begin{split}
   r^* &= \frac{G}{1 + N^{{\scriptscriptstyle(C)}}_C-N^{{\scriptscriptstyle(D)}}_C} \\
   &= \frac{G}{1+\Delta N_C}.
   \end{split}
   \label{rG-limit}
\end{equation}
in the limit of low noise $K$ (i.e., the deterministic approach), where $N^{{\scriptscriptstyle(C)}}_C$ is the number of cooperative players around a cooperator and $N^{{\scriptscriptstyle(D)}}_C$ the number of cooperators around a defector and both are randomly chosen from the whole population with  $0\leqslant N^{{\scriptscriptstyle(j)}}_C\leqslant G-1$ for $j = C,D$. 
Equation~\ref{rG-limit}, together with the conditions on $N^{{\scriptscriptstyle(C)}}_C$ and $N^{{\scriptscriptstyle(D)}}_C$, gives rise to a series of transition values responsible for the survival of cooperation in different microscopic configurations.  
Since the neighborhood is randomly chosen from the entire population, the probability of having a $C$ as a neighbor is the same for both interacting sites, irrespective of their strategies.
Therefore, we will have pairs of microscopic configurations with $\Delta N_C =\pm \,n$, with $n \in \{1, 2,\dots , G - 1\}$ which occur with the same probability. The outcome will consequently be determined by the $\Delta N_C = 0$ configurations, which only favor cooperators when $r > G$.
As a result, there will be more situations that are favorable than are detrimental to cooperators in this region, and consequently, it will be more likely for them to survive. 

In a different implementation of the random neighborhood, we considered that both interacting sites are fixed as participants in each other's group. In this case, even in the region $r>G$, the defector will always have at least one cooperator in their group, resulting in the extinction of cooperation.
The random PGG case will be discussed later in section~\ref{sec_pgg}.
\subsubsection*{Evolution on regular lattices} 

\begin{figure}[t]
\centering
{\includegraphics[width=\linewidth]{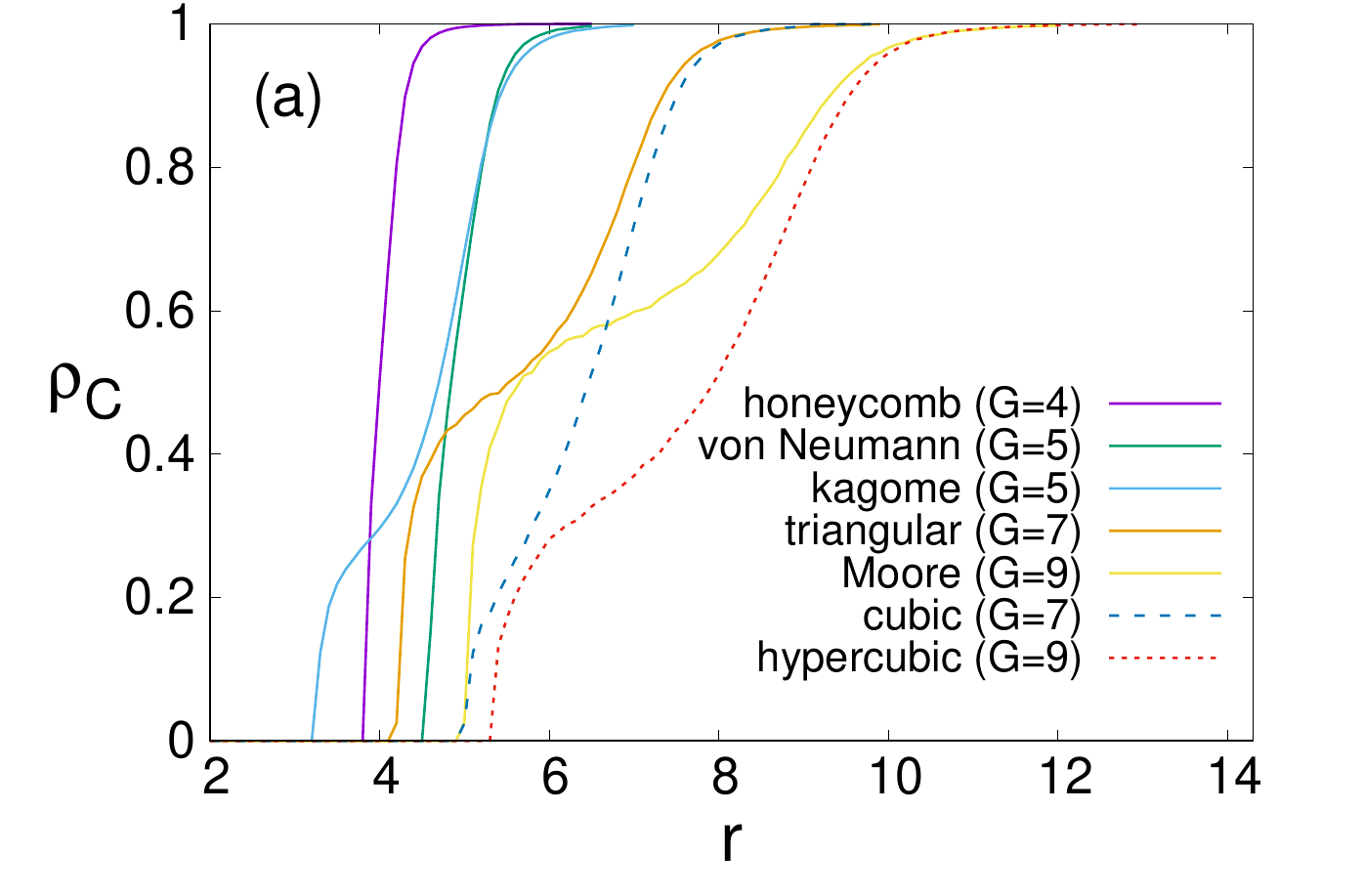}}
\includegraphics[width=\linewidth]{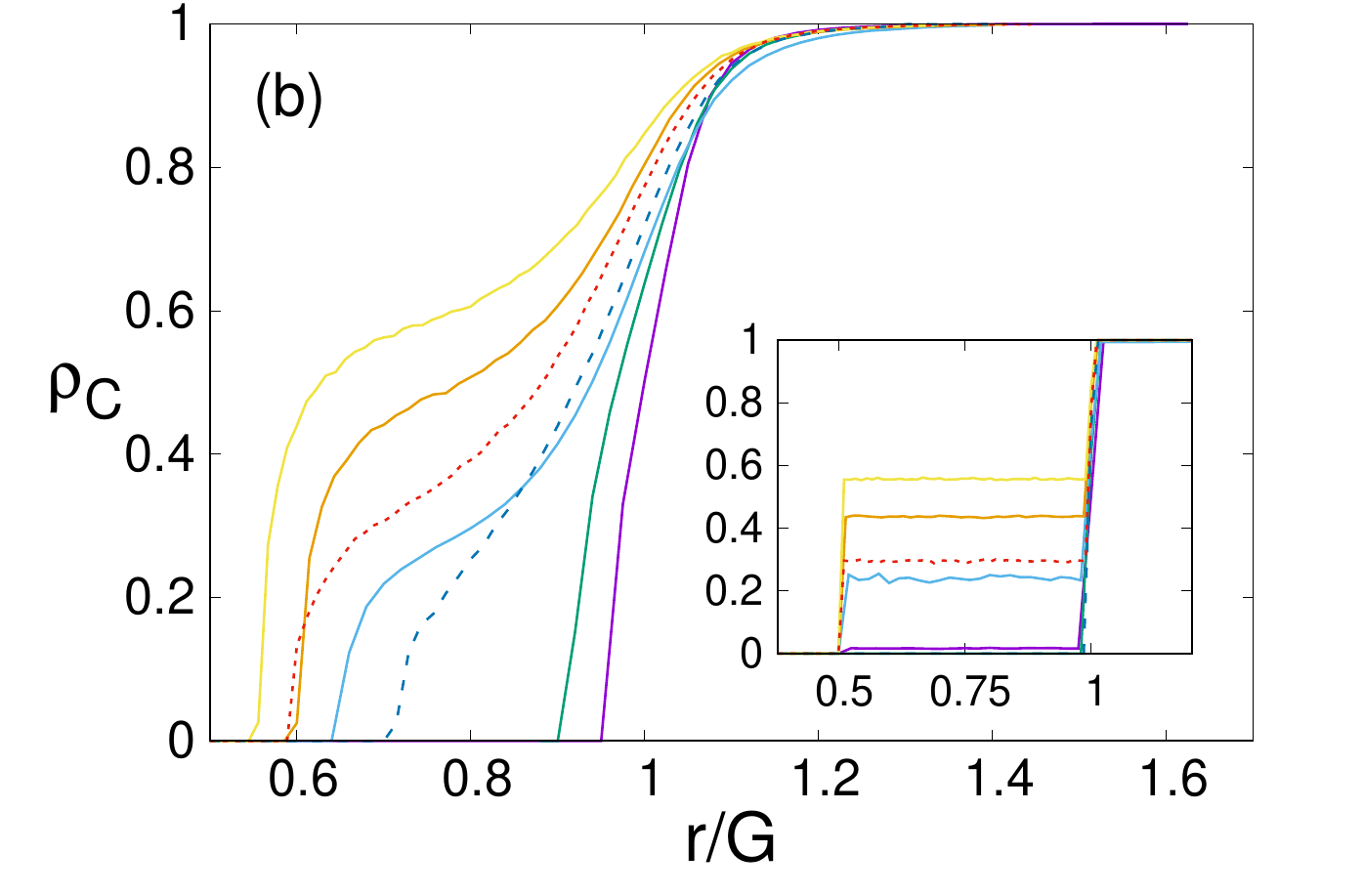}

\caption{ (a) Asymptotic  fraction of cooperators, $\rho_C$, as a function of $r$ for the Focal Public Goods Game (FPGG). (b) shows cooperator density as a function of the normalized multiplicative factor $r/G$ in the FPGG with $K=0.1$. The inset shows the analogous results with a low noise value, $K=0.001$. The normalized data indicates that an increasing number of neighbors favors the settlement of cooperative behavior. All topologies with common neighbors between two connected players are the most beneficial for cooperation. Even more, with the increase in common neighbors, we see higher cooperation density in the coexistence region. }
\label{norm}
\end{figure}

Now, we investigate the effect of the different lattices described in Sec.~\ref{sec.topologies} on the final fraction of cooperators, $\rho_C$. 
First, we present the raw data with $K=0.1$ in Fig.~\ref{norm} (a), showing that a group with more connections needs a greater return to compensate all group members. This results in a higher $r$ value to sustain cooperation for the more connected lattices.
The general behavior of $\rho_C$ as a function of the normalized synergistic factor, $r/G$, for the FPGG is presented in Fig.~\ref{norm} (b) for the studied topologies.
The first thing to notice is that, in all cases, $\rho_C\to 1$ around the same value of $r/G$. 
When looking only at the two-dimensional lattice arrangements in Fig.~\ref{norm}~(b) (disregarding the cubic and 4D hypercubic lattices), we can see a very interesting behavior: for all lattices, there is a coexistence region. In this region, the cooperation level is higher for the most connected topologies (higher $G$). That is, the Moore neighborhood ($G=9$) has a higher cooperation level than the triangular ($G=7$), which in turn has more cooperation than the kagome and square lattices, both with $(G=5)$. Lastly, the honeycomb lattice ($G=4$) is the one with the lowest cooperation level. Again, this very interesting hierarchy of the 2D-lattices is only visible when using the normalization $r/G$. 
The fact that the cubic and 4D lattices do not follow this trend reinforces that a change in dimensionality is not only a matter of increasing the number of a given site's connections, but rather it is a non-trivial topological change.
Besides this point, we can see that lattices with more neighbors in common between two connected players are the most beneficial to cooperation. There is a higher cooperation fraction in the kagome than in the von Neumann neighborhood, both sharing the same number of agents in each group, $G=5$. 

Next, we investigate the configurations of strategies that allow the survival of cooperators. 
We present in the inset of Fig.~\ref{norm} (b) the final cooperation density as a function of the normalized parameter for different lattices in the case where $K=0.001$, i.e., very close to the deterministic case. Comparing it with the possible values of $r/G$ from eq.~\ref{rG-limit}  we can see which microscopic configurations are the most frequent in the system's dynamics.
The figure shows that changes in the cooperator density occur only for values of $r^*/G=1/2$ and $1$. 
This shows that situations in which a cooperator has $1$ or $2$ more $C$ neighbors in its group (including themselves) than the competing defector are the most important for the survival of cooperation, independently of the topology.
Such configurations can be traced to cooperators at cluster boundaries trying to survive the invasion by a defector (see Appendix~\ref{app}). 
In the region $r/G>1$,  all configurations in which a cooperator has more $C$ neighbors than a defector will favor the former. Therefore cooperation will thrive for any topology.
It is important to note that differently from the random neighborhood case, this situation exists due to clustering and not because of the presence of the focal cooperator in their group, since now they are also a member of the interacting defector's group.
Now, for $r/G<1$, there will be situations in which a cooperator will be invaded even with more $C$ neighbors than the defector.  
Depending on the topology, this can either mean the extinction of cooperation or a region of coexistence between the two strategies. 
\begin{figure}[t]
\begin{center}

\includegraphics[width=.24\columnwidth]{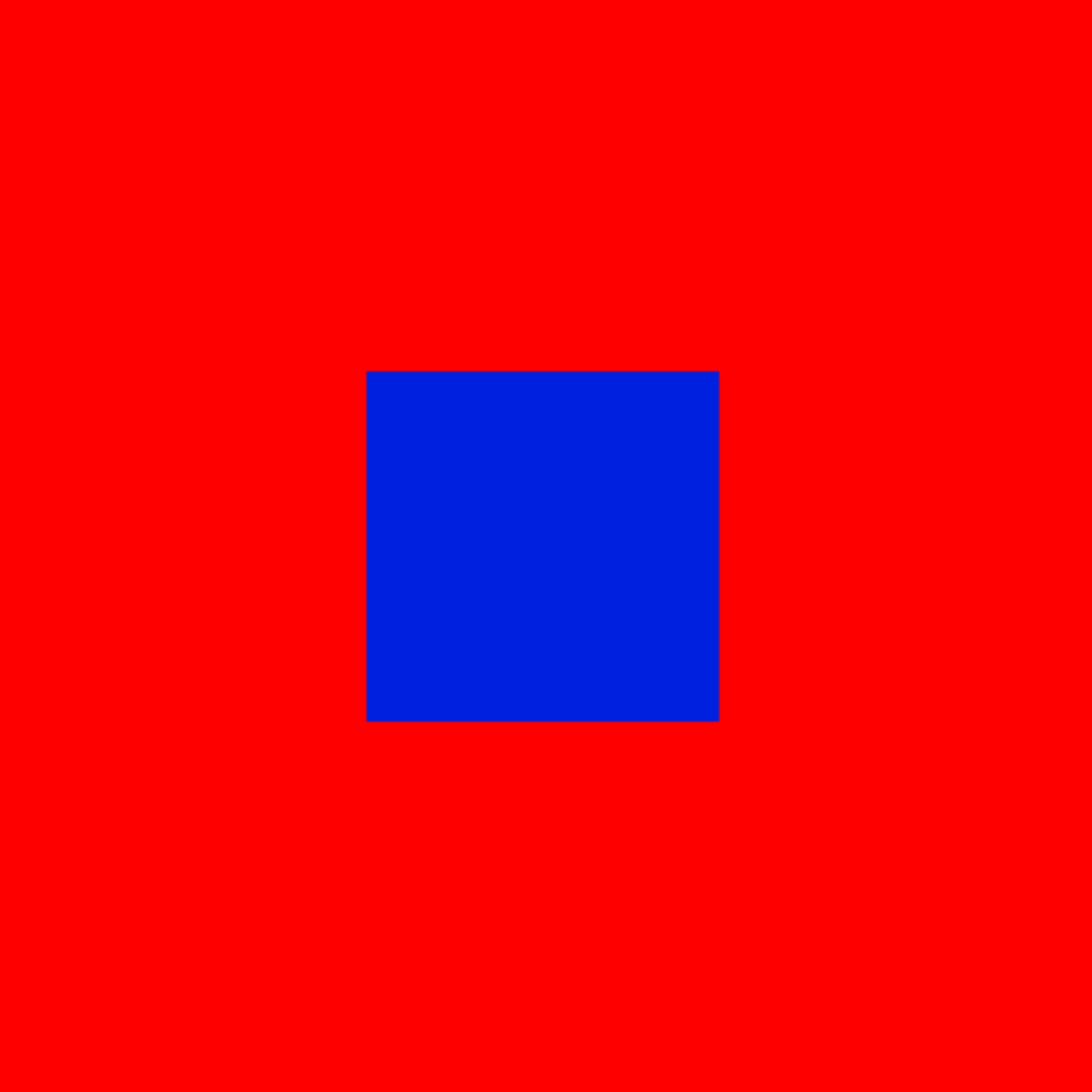}
\hspace{-.15cm}
\includegraphics[width=.24\columnwidth]{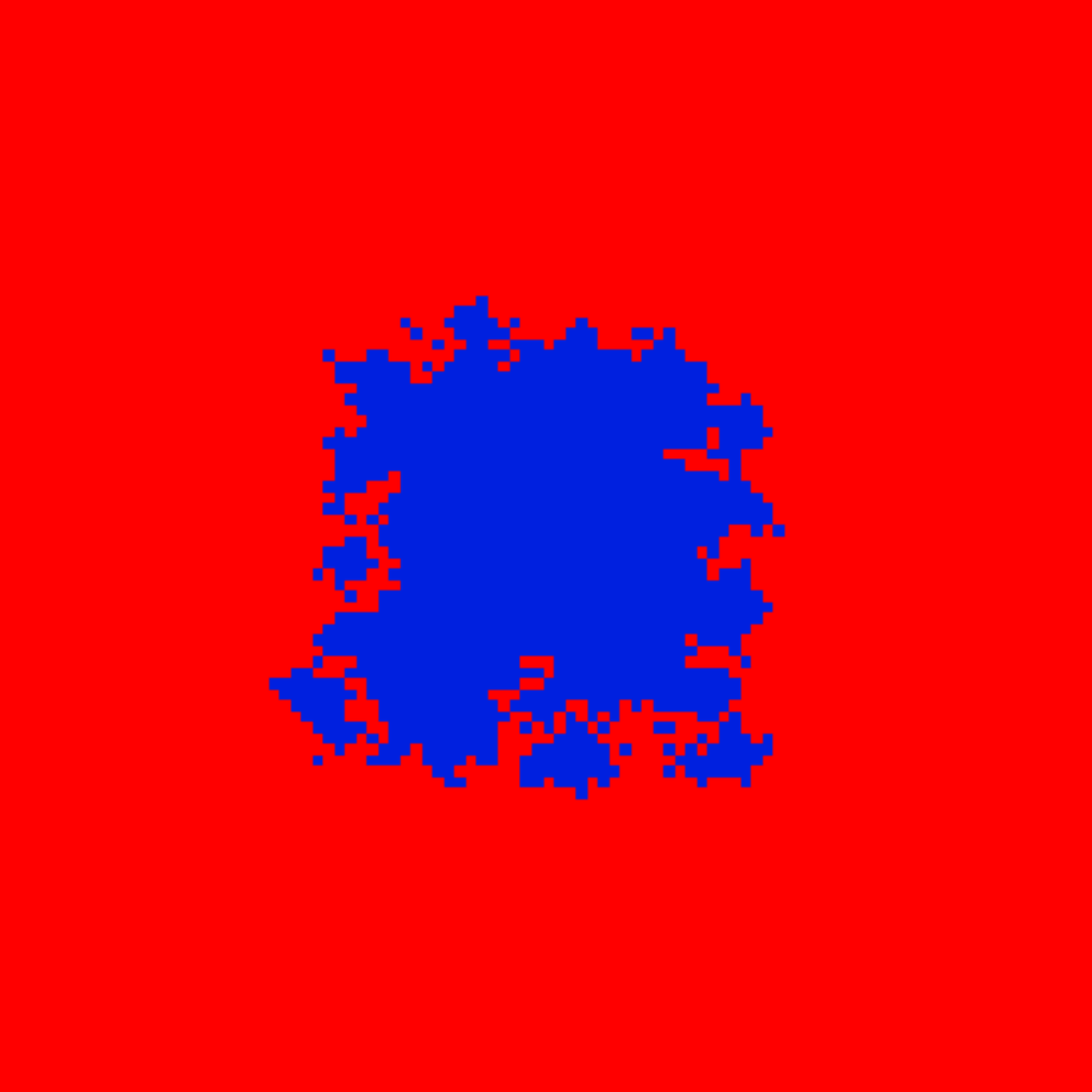}
\hspace{-.15cm}
\includegraphics[width=.24\columnwidth]{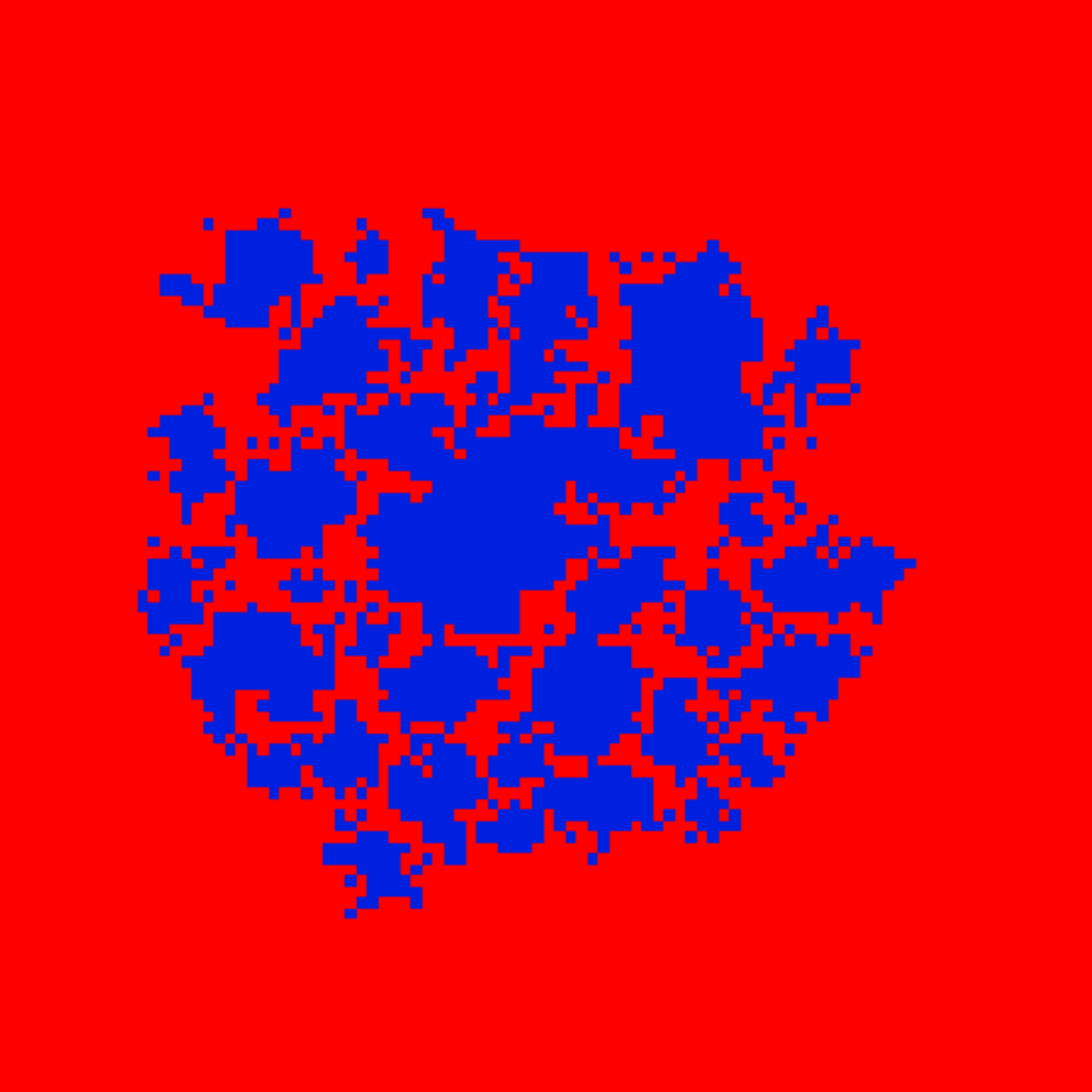}
\hspace{-.15cm}
\includegraphics[width=.24\columnwidth]{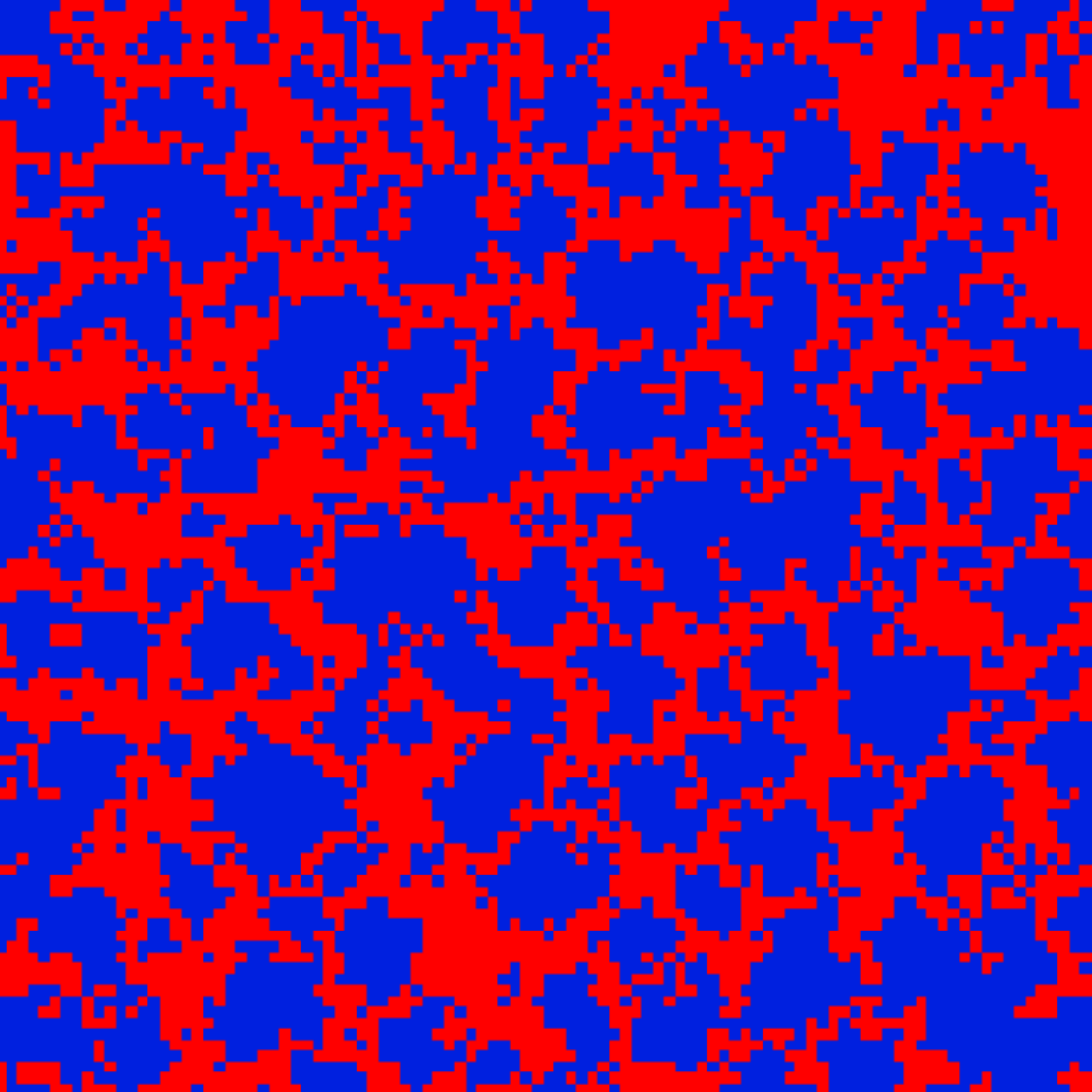}

\vspace{0.01cm}
\includegraphics[width=.24\columnwidth]{Moore11.pdf}
\hspace{-.15cm}
\includegraphics[width=.24\columnwidth]{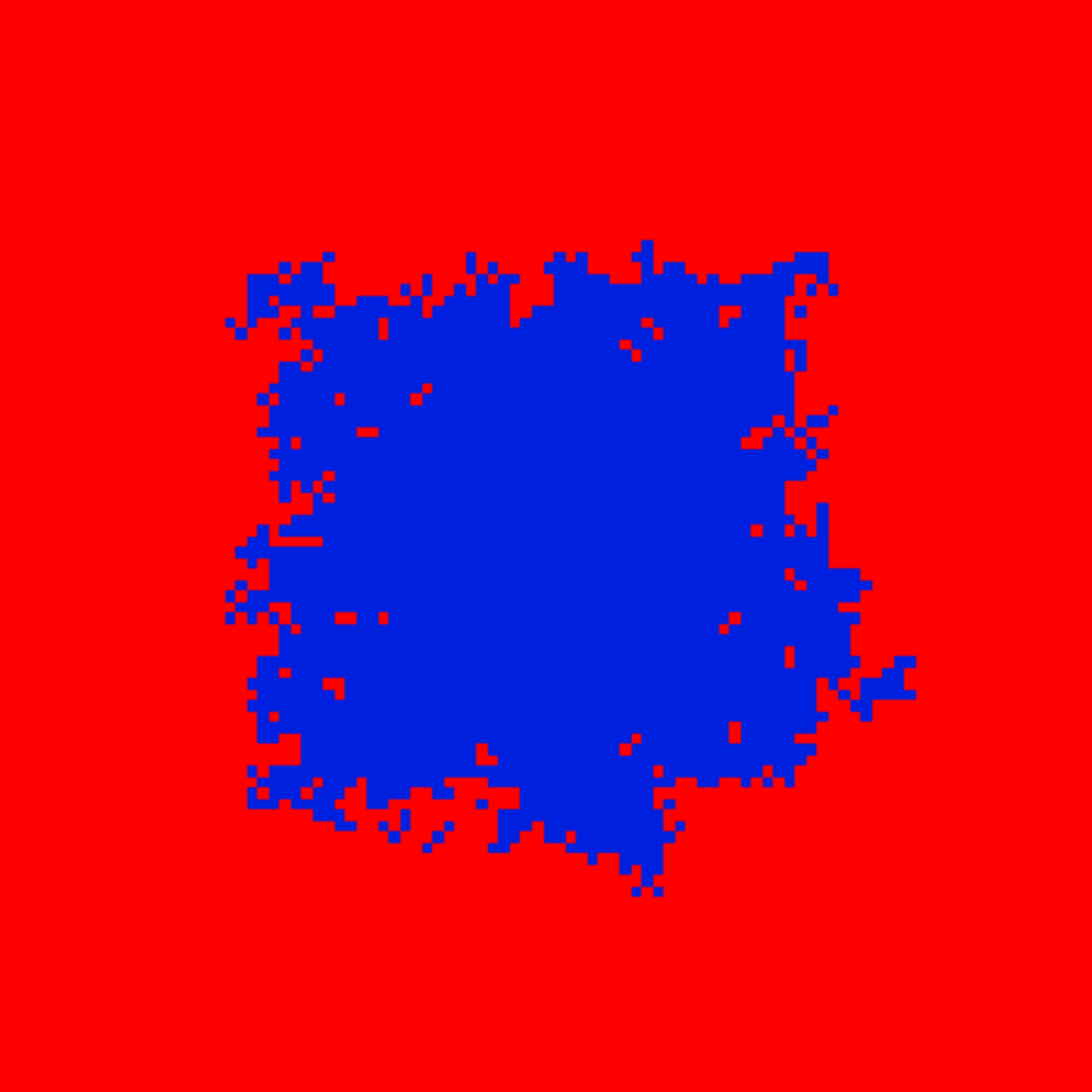}
\hspace{-.15cm}
\includegraphics[width=.24\columnwidth]{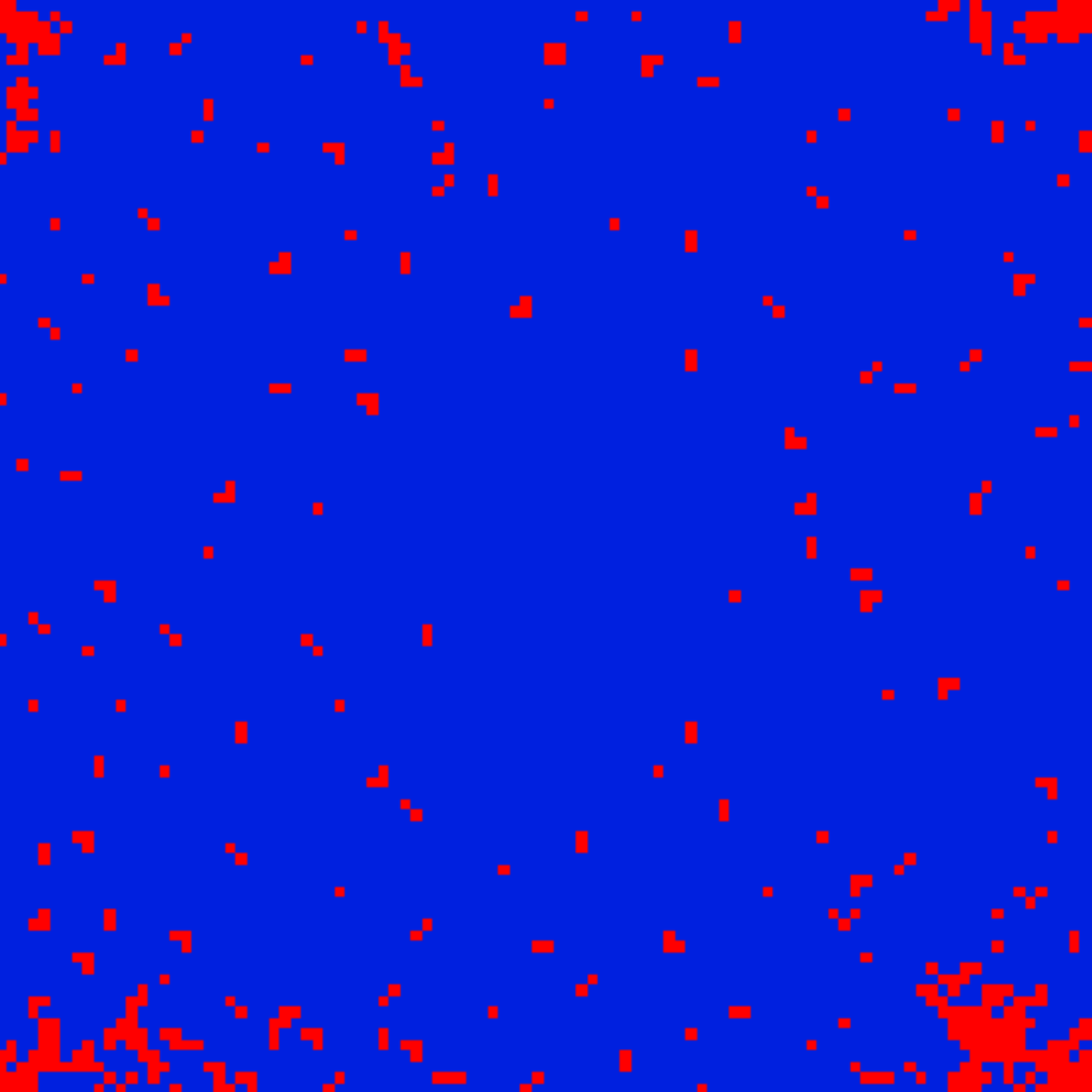}
\hspace{-.15cm}
\includegraphics[width=.24\columnwidth]{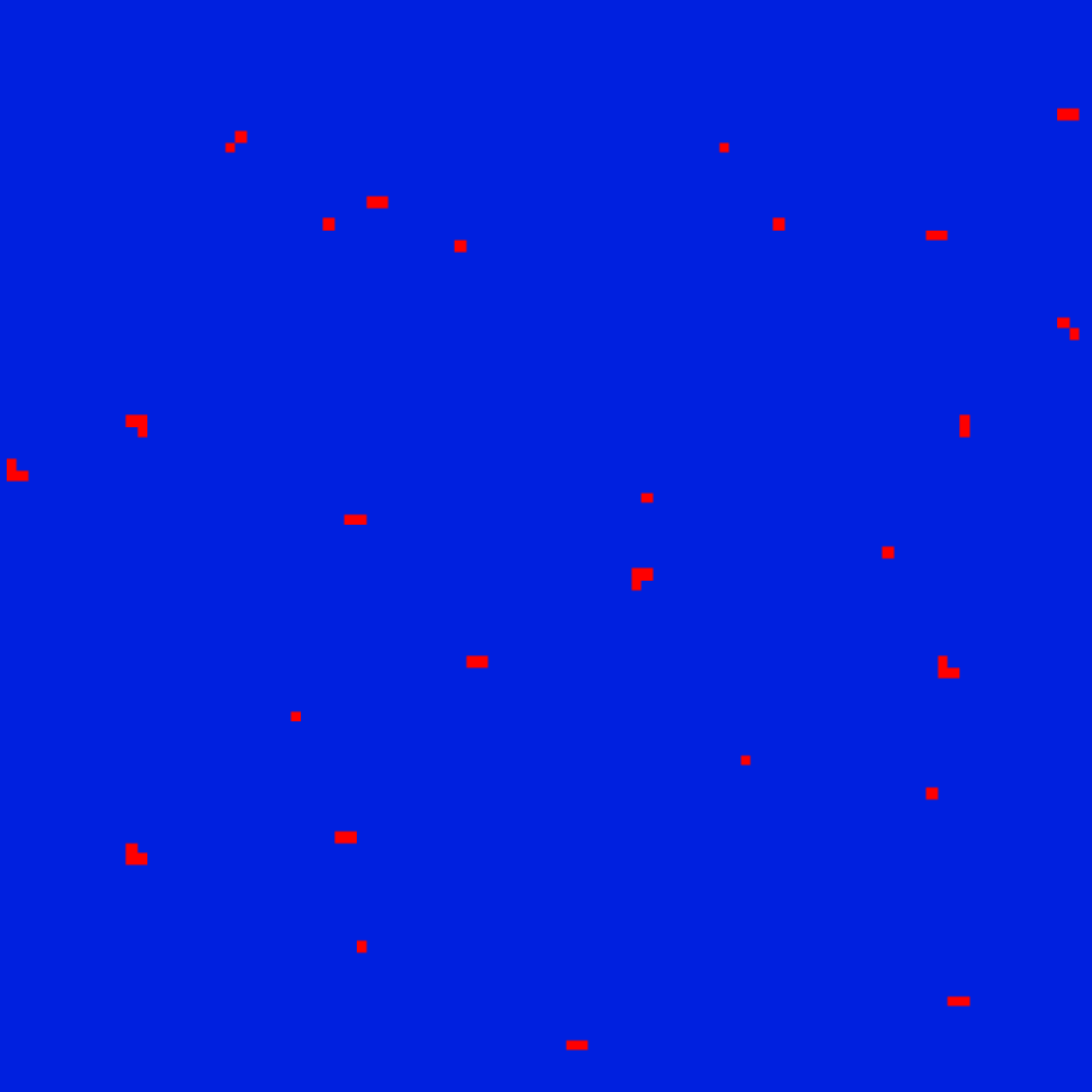}
    \caption{Time evolution of a cooperator cluster  (blue) in the sea of defectors (red) on the square lattice with the  Moore neighborhood for $K=0.001$ ($t=0$, $25$, $100$, $1000$). The top row illustrates the coexistence region ($r/G=0.98$) where, despite the $C$ cluster expansion, defectors survive by infiltration. The bottom row shows the dominance region ($r/G=1.01$), where this invasion is impossible, leading defectors to extinction. }
    \label{corte}
    \end{center}
\end{figure}
For the von Neumann neighborhood, this situation is enough to extinguish cooperation, as seen in the inset of Fig.~\ref{norm}~(b). 
For the triangular lattice and Moore neighborhood, a cluster of cooperators can grow while being invaded by a line of defectors in the region $1/2<r/G<1$. Figure~\ref{corte} shows snapshots of the time evolution in the Moore neighborhood for both regions, where the possibility of penetrating the cooperator cluster ensures the coexistence of both strategies. The kagome lattice presents similar dynamics, as discussed in the Appendix~\ref{app}.

The honeycomb lattice is a peculiar case, having the smallest number of connections ($G=4$). Because of that, there are very few possibilities for a cluster of $C$'s to expand or to be invaded. Thus, if we start with a random initial condition, the cooperator density will decrease until only clusters formed in the initial condition survive. In this scenario, a cluster of $C$'s can neither expand nor be invaded if $1/2<r/G<1$, explaining the coexistence region and the low density of cooperation shown in the inset of Fig.~\ref{norm}~(b).
As a general rule, more connected lattices allow more links among cooperators due to clustering. However, since this also happens between defectors and cooperators, it is not obvious that the increase in the number of connections would favor cooperation.
Nevertheless, the number of microscopic configurations that give a fixed transition value in eq.~\ref{rG-limit} and benefit cooperation increases with the number of connections, despite all lattices sharing the same $r/G$ transitions values (see Appendix~\ref{app}). 
Therefore, cooperators survive in a greater number of configurations on highly connected lattices.

\subsection{Public Goods Game}
\label{sec_pgg}

\subsubsection*{Evolution on regular lattices}

\begin{figure}[t]
\centering
\includegraphics[width=\linewidth]{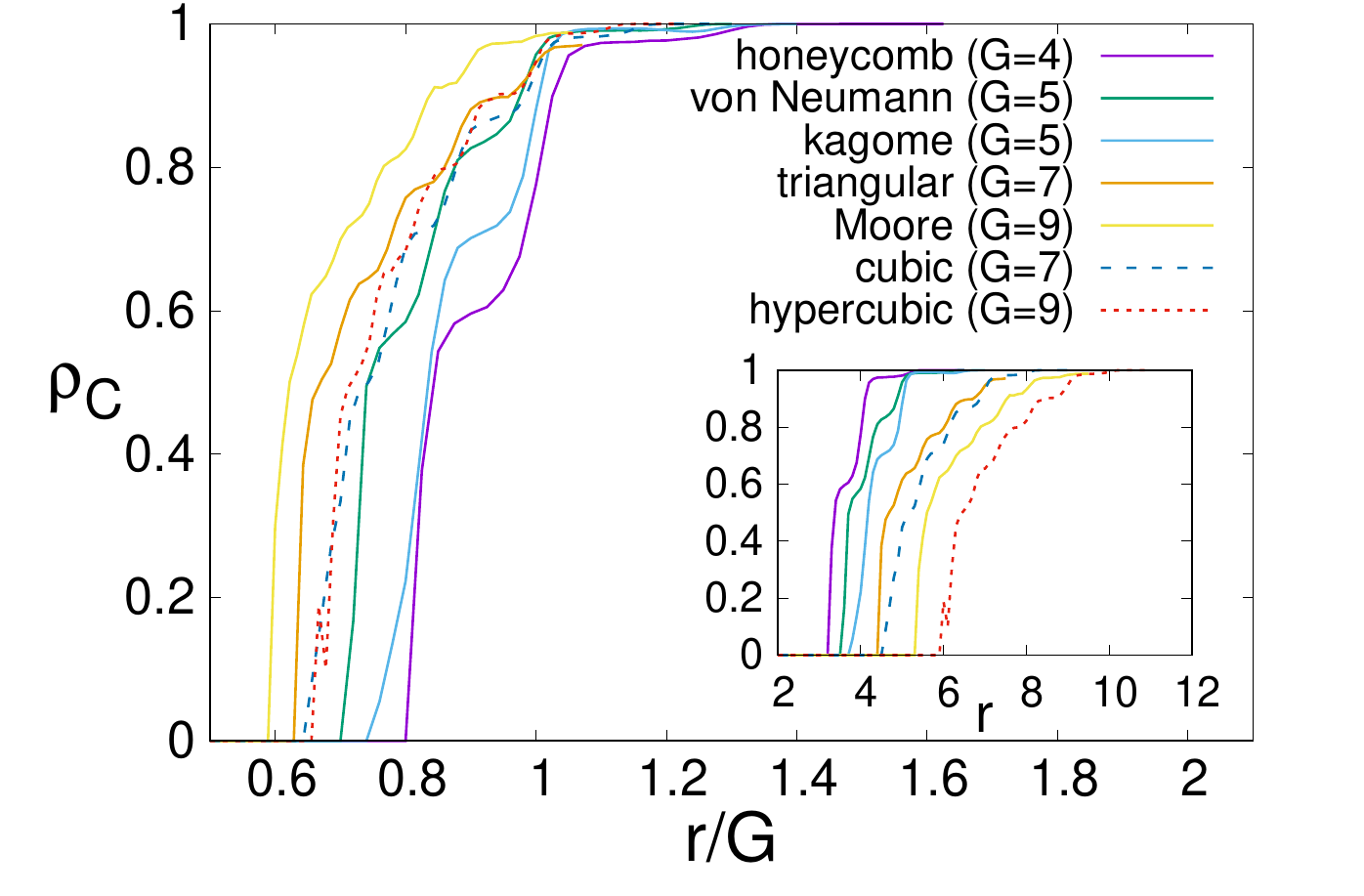}
\caption{ Final fraction of cooperators, $\rho_C$, as a function of $r/G$ for the classic Public Goods Game (PGG), with $K=0.1$. The inset shows the same data without normalizing $r$ by the group size on each lattice, $G$.  Normalized scales suggest a trend in both PGG games, where the presence of an increasing number of neighbors favors the settlement of cooperative behavior.} 
\label{normpgg}
\end{figure}

Now we will investigate the effects of connections for the classic PGG, where now agents participate in one game centered on themselves and one game centered on each of its neighbors (remember that the total number of neighbors, and thus games played, will be different for each topology). In Fig.~\ref{normpgg}, we present the final fraction of cooperators as a function of $r/G$.  
The first thing to note is that in general, the fraction of cooperation will be higher in this case than in the analogous FPGG for the same value of $r/G$.  
This is expected~\cite{szolnoki09} since a cooperator now participates in $G$ groups, earning  $G$ self-contributions. Therefore the positive effect of spatial reciprocity is amplified as previously discussed in Sec.~\ref{sec.FPGG}.

Regarding the different topologies, we see that the previous ordering remains unchanged: cooperation fraction increases with the increasing number of neighbors on two-dimensional lattices. 
Similar to the FPGG, we observe that the cubic and $4D$ lattices do not follow this pattern. 
It's interesting to see that the increase in connections from the FPGG to the PGG allows clusters in the honeycomb lattice to expand, resulting in high levels of cooperation in the coexistence phase.

\begin{figure}[t]
    \centering
    \includegraphics[width=\linewidth]{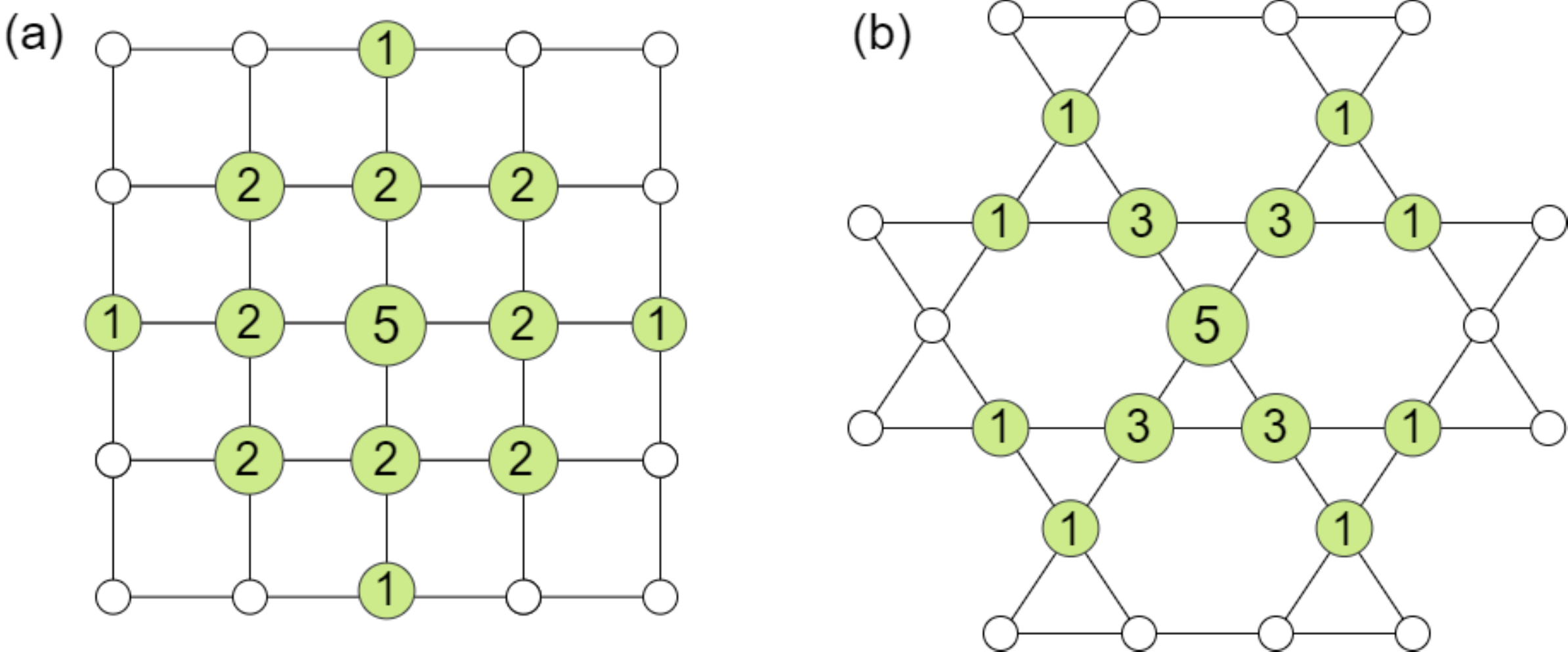}
    \caption{Illustration of the classic PGG viewed as a Focal PGG for (a) the square lattice with von Neumann neighborhood and (b) kagome lattice. For this approach, each neighbor is taken into account with a different weight in the focal player's payoff, since it participates in a different number of groups with the focal site, depending on lattice structure. It is important to notice that despite the group sizes highlighted shown above, the focal player can only interact (compare payoffs to decide whether to change strategy) with one of the first neighbors.
    }
    \label{pgg2}
\end{figure}

Contrary to what was found in the FPGG, the square lattice with the von Neumann neighborhood has a higher cooperation fraction than the kagome lattice in the PGG. 
To analyze this change of behavior, we propose an approach similar to the one applied to the FPGG: the classic PGG can be formulated as a FPGG with an extended interaction region for the calculation of payoffs. Now, each player is taken into account with a different weight in the focal player's payoff since each neighbor will play (and contribute) to a different number of games with the focal player.
The payoff's calculation for any topology works as follows: a focal cooperator will always have $G$ contributions from themselves (they participate in $G$ different games); therefore, the focal player is taken into account with a weight equal to $G$. 
All $G-1$ first  neighbors have a weight that results from the sum of the following contributions: one (from the game centered on themselves) plus one (from 
the game centered on the focal player) plus the number of links with other first neighbors (common neighbors with the focal player).
All other players in the population (which are not first neighbors with the focal player) have a weight equal to the number of their connections to the first neighbors. 
Figure~\ref{pgg2} illustrates this approach for the von Neumann neighborhood and kagome lattice. The focal site's weight is five because it contributes to all five groups. Similarly, all players with weight two participate in two groups of which the focal player is also a member, etc.
Now the payoffs of a cooperator on the square (von Neumann neighborhood) and the kagome lattice are, respectively,
\begin{align}
    \Pi_{C}^S&= G\left(\frac{r c}{G}\right)+2 N_{C_2}^{{\scriptscriptstyle(C)}} \left(\frac{r c}{G}\right)+N_{C_1}^{{\scriptscriptstyle(C)}} \left(\frac{r c}{G}\right)-G c \label{eq_square}\\
    \Pi_{C}^K&=G\left(\frac{r c}{G}\right)+3 N_{C_3}^{{\scriptscriptstyle(C)}} \left(\frac{r c}{G}\right)+N_{C_1}^{{\scriptscriptstyle(C)}} \left(\frac{r c}{G}\right)-G c \label{eq_kagome}
\end{align}
where $N_{C_i}^{{\scriptscriptstyle(C)}}$ is the number of cooperators with weight $i$ for that given topology. 
From an inspection of Fig.~\ref{pgg2}, we see that $0\leqslant N_{C_2}^{{\scriptscriptstyle(C)}}\leqslant 8$ and $0\leqslant N_{C_1}^{{\scriptscriptstyle(C)}}\leqslant 4$ for the von Neumann neighborhood, and $0\leqslant N_{C_3}^{{\scriptscriptstyle(C)}}\leqslant 4$ and $0\leqslant N_{C_1}^{{\scriptscriptstyle(C)}}\leqslant8$ for the kagome lattice.
Although the spatial distribution of cooperators inside the group was not important for the FPGG, here this is not the case:  the cooperators' locations are relevant since common neighbors between players increase the weight of a neighboring site. Defector payoffs are calculated in a manner analogous to the expressions above. 
With this, the transition values for $r$ (analogous to eq.~\ref{rG-limit}),
 given for example for the von Neumann case by
\begin{equation}
\begin{split}
    r^{*} &=\frac{G^2}{G+2\left(N_{C_2}^{\scriptscriptstyle(C)}-N_{C_2}^{\scriptscriptstyle(D)}\right)+\left(N_{C_{1}}^{\scriptscriptstyle(C)}-N_{C_1}^{\scriptscriptstyle(D)}\right)} \\
    &= \frac{G^2}{G+2\,\Delta N_{C_2}+\Delta N_{C_1}},
    \end{split}
      \label{eq.trans_PGG}
\end{equation}
now depend nonlinearly on $G$.
Therefore, different topologies present distinct normalized transition values ($r^{*}/G$), in contrast to the FPGG. 
Another important fact is that highly connected lattices possess more transition values than lattices with fewer connections. This shows that the increase in connections in the PGG game allows more scenarios in which cooperators have more $C$ neighbors than defectors. We remark that although payoffs are calculated in an extended region, strategy flips occur only between first neighbors, as in the FPGG.

\begin{figure}[t]
    \centering
   \includegraphics[width=\linewidth]{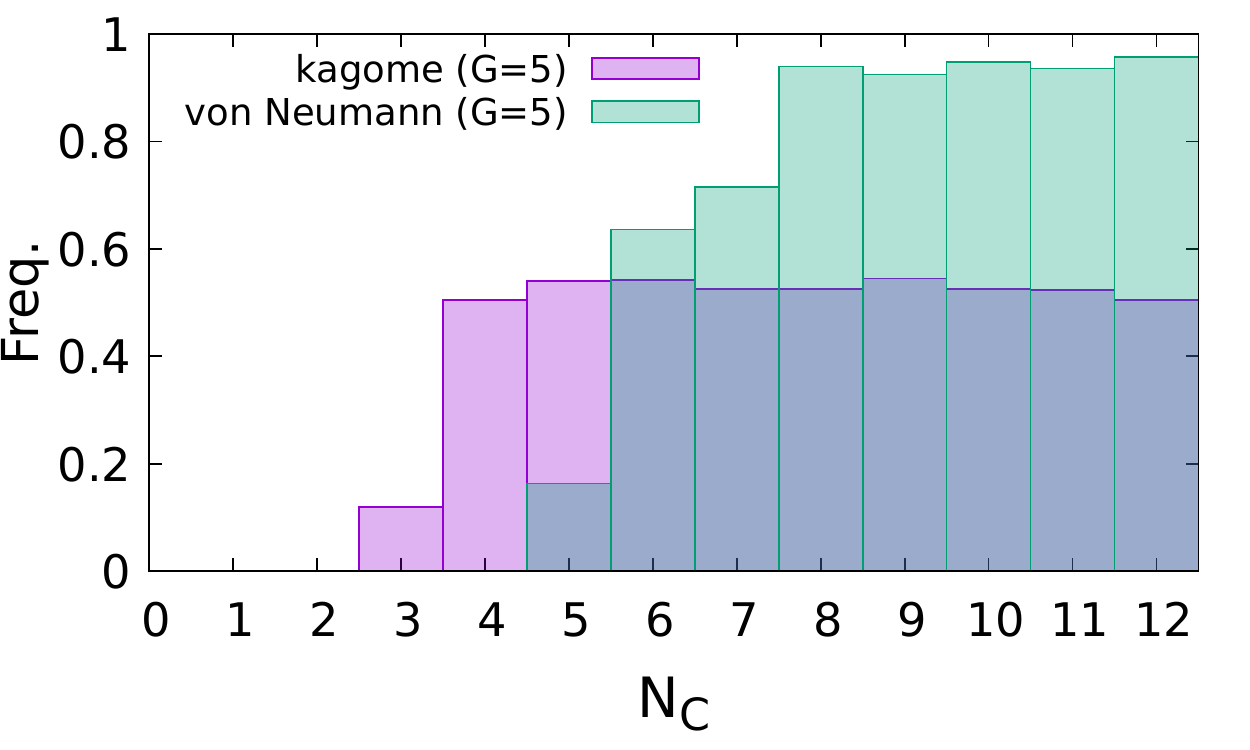}
        \caption{Fraction of simulations ($n=1000$) starting with a cooperator cluster of size $1+N_C$ in a sea of defectors in which cooperation survives for the kagome and square lattice with von Neumann neighborhood. The initial cluster is created by sequentially distributing $N_C$ cooperators in layers around a seed cooperator.
        }
    \label{dif}
\end{figure}

Since the size of a group is larger for the PGG than for the FPGG, it is harder for cooperators to become isolated in the former. Therefore, it is interesting to see how cluster size influences the survival of cooperation.
With this purpose in mind, we simulated the dynamics of a single cluster of cooperators in a sea of defectors. We assembled a single initial seed cooperator with a varying number of $C$ neighbors, $N_C$, for a fixed value of $r$ ($r/G = 0.8$) and $K = 0.001$ (low noise). The cluster is formed by sequentially distributing the cooperators in layers: initially in the first neighbors, then in the other sites with high weights, and finally in sites with lower weights. The fraction of simulations in which cooperation survives with different initial cluster sizes is shown in Fig.~\ref{dif}.
When cooperation survives it always reaches a constant density,  $\rho_C\approx 0.54$ and $\rho_C\approx 0.22$, for the von Neumann neighborhood and the kagome lattice, respectively (note that these are essentially the same densities attained from the random initial condition with $50\%$ cooperators, see Fig.~\ref{normpgg}).
Comparing both topologies we see that for $N_C\leqslant 2$, cooperation is always extinct in both of them. 
For $ 3 \leqslant N_C \leqslant 5$, cooperation is extinct more frequently in the von Neumann neighborhood than in the kagome lattice, whereas for $ 6 \leqslant N_C \leqslant 12$, the converse takes place. 
%
In both topologies, the same pattern occurs: cooperation is more easily established with an increase in $N_C$ until the first layer of high-weighted value sites is occupied. After that, increases in the number of  $C$ neighbors are not beneficial to cooperation in general.
This can be understood by noticing that if a defector occupies a high-weight value site, many of the cooperators from the initial cluster will also be their neighbors, thus allowing more invasions to occur. Since for lower $N_C$ more of these defectors are present, it is more likely for the cluster to be invaded and destroyed.
On the other hand,  all the $C$'s at the border could already invade the sea of defectors for $r/G > r^*/G =  5/7$. Therefore an increase in $N_C$ beyond the point at which the first layer with all high-weighted value sites is occupied does not change the behavior of the cluster.
For example, if $N_C=8$ for the von Neumann case, a $C$ at the border can invade a defector in the location where a new $C$ with weight $1$ would be. Thus, the case $N_C=9$ is nothing more than the case $N_C=8$ at a later time in the evolution of the system (the analogous situation happens for the kagome lattice when $N_C=4$).
Thus, for the kagome lattice, both small and large clusters ($N_C\geqslant 4$) are equally likely to survive, while for the von Neumann neighborhood, larger clusters are necessary.

For the initially random condition with $50\%$ cooperators, we saw in Fig.~\ref{normpgg} that there was a higher density of cooperation with the von Neumann neighborhood than in the kagome lattice. 
This is because the initial density of $C$s was high enough for clusters with a high $N_C$ to exist and expand, while small ones disappear in the first generations. And, as noted already in Nowak's introduction of spatial structure in evolutionary games~\cite{Nowak1992a}, the asymptotic density is almost completely independent of the initial condition.
Additionally, if we set the initial random condition with $\rho_C = 0.1$, large clusters become unlikely. Now cooperation cannot survive in the von Neumann neighborhood while it is still possible in the kagome lattice, corroborating our arguments.

\subsubsection*{Random neighborhood}
\label{sec_rand_PGG}

Recalling the random neighborhood discussed for the FPGG, we saw that cooperation only survived for $r > G$.
Equation~\ref{eq.trans_PGG} refers to the transition values for the von Neumann neighborhood, but its form is similar for all lattices. 
Since we choose the neighborhood randomly from the entire population, the chance of having a $C$ in the group is the same for both interacting sites, regardless of their strategies. 
This means that there will be the same number of situations that are favorable and prejudicial to a cooperator with equal probability, considering $\Delta N_{C_1} \neq 0$  and $\Delta N_{C_2}  \neq 0 $ (see Sec.~\ref{sec.random}).
However, if $r>G$, all situations involving $\Delta N_{C_1} = \Delta N_{C_2}  = 0 $ will favor  cooperators, making the total number of microscopic configurations that favor cooperation greater than the number of those that don't.  

Another important aspect of the classic PGG is that a cooperator's payoff has $G$ contributions from themselves due to all $G$ groups in which they participate. 
This high gain is then able to compensate the exploitation by defectors for some configurations, even  when both $\Delta N_{C_1} < 0$  and $\Delta N_{C_2}  < 0 $. 
Consequently, transition values $r^* > G$ now become possible in contrast with the FPGG, in which cooperation couldn't survive for any parameter value if the number of cooperators was equal to or less than the number of defectors in a group (see eq.~\ref{rG-limit}).
Therefore, if we consider that both interacting sites are fixed in each other's group, there exists a finite value of $r$, with $r>G$,  above which cooperation can be sustained.

\subsection{Prisoner's Dilemma}
\label{pd_sec}

We also analyze the Prisoner's Dilemma game on the same topologies.
The most general formulation of the PD game demands that $T>R>P>S$ and $2R>T+S$, which allow many different possible parametrizations for the payoffs. 
A commonly used one is to define a benefit $\beta$  and a cost $\gamma$, both positive, and set $T = \beta$, $R = \beta - \gamma$, $P = 0$ and $S = - \gamma$. In this case, it is possible to directly map the PD into the FPGG by equating the payoffs for cooperators (defectors) of the two games; the case without self-interaction and with the contribution parameter of the FPGG fixed, $ c=1$, was discussed in~\cite{Hauert2003}.  The case with self-interaction presents some subtleties, which we describe here. Resorting to the FPGG payoffs, eqs.~\ref{payD} and \ref{payC}, and to the PD payoffs
\begin{align}
    \Pi_{D,PD} &= N^{{\scriptscriptstyle(D)}}_C \,T + (G-N^{{\scriptscriptstyle(D)}}_C)P\\
    \Pi_{C,PD} &= (1+N^{{\scriptscriptstyle(C)}}_C) R + \left[G-(N^{{\scriptscriptstyle(C)}}_C+1)\right] S,
\end{align}

we can write
\begin{align}
\begin{aligned}
T&=\frac{r c}{G} \equiv \beta \\
R&=\frac{r c}{G}-\frac{c}{G}=\beta-\gamma \\
P&=0 \\
S&=-\frac{c}{G}=-\gamma, 
\end{aligned}
\end{align}
from which we obtain  
\begin{align}
\label{eq_rmap}
    r &= \frac{\beta}{\gamma} \\
\label{eq_rmap2}
    c &= \gamma\, G
\end{align}
for the  corresponding FPGG; it is necessary to have $r>1$ to obey the PD's payoff hierarchy (to preserve $\beta>\gamma$). 
Care must be taken when performing this mapping, especially when comparing different topologies,  because now the value of the contribution, $c$, depends on the group size, but it is usually fixed at unity when simulations of the FPGG are performed. This is not a problem when analyzing the games on only one topology, since,  as seen in Sec.~\ref{model}, changing the contribution can be viewed as a rescaling of the noise from $K$ to $K'=K/c$ and we can map both games onto one another on a fixed lattice, but the result will be related to the rescaled noise $K'$ in the FPGG.
Of course, if the contribution value is corrected according to  eq.~\ref{eq_rmap2} in the FPGG, there is no need to take into consideration the change in noise.

\begin{figure}[t]
    \centering
{\includegraphics[width=\linewidth]{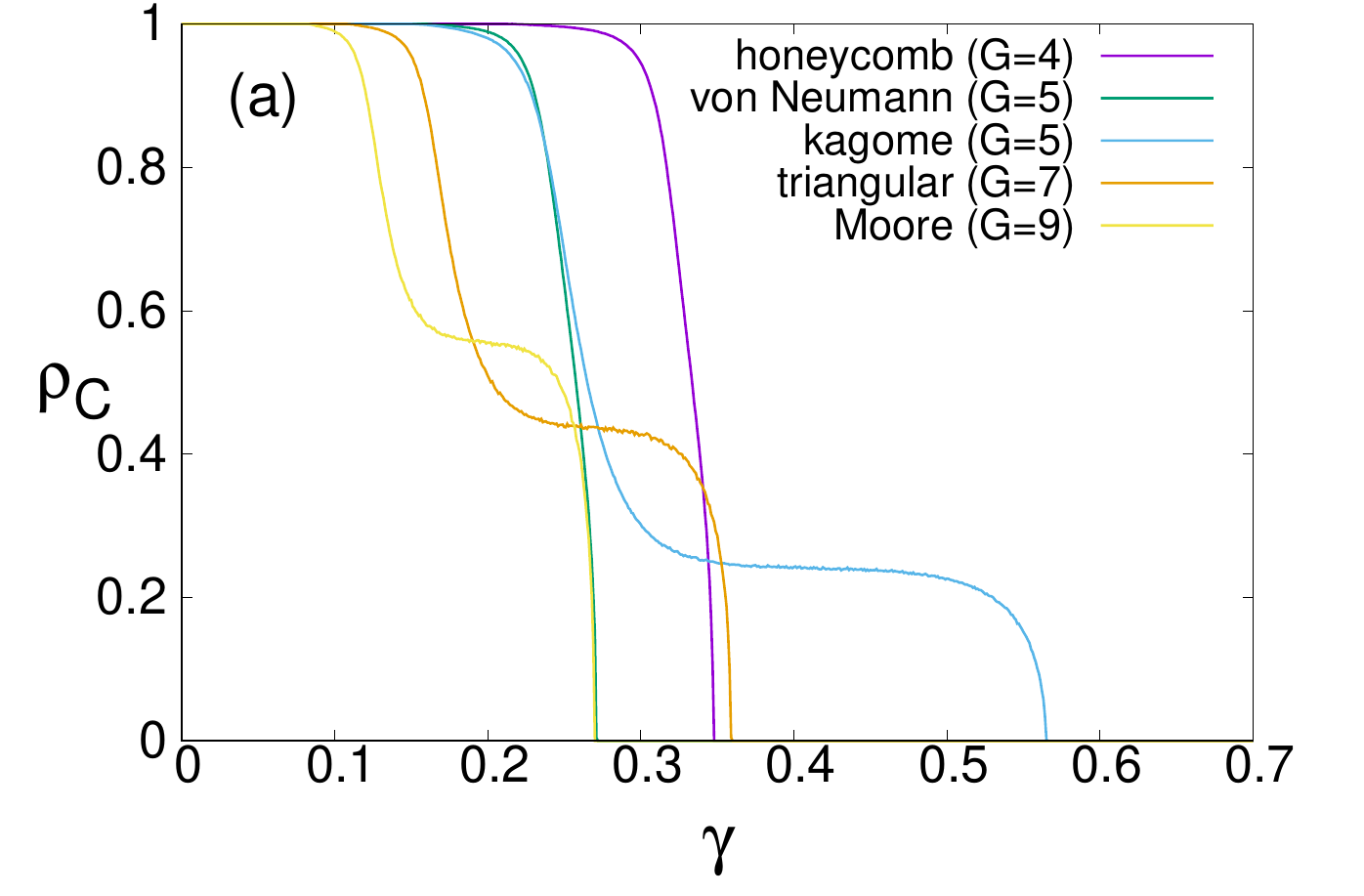}}
{\includegraphics[width=\columnwidth]{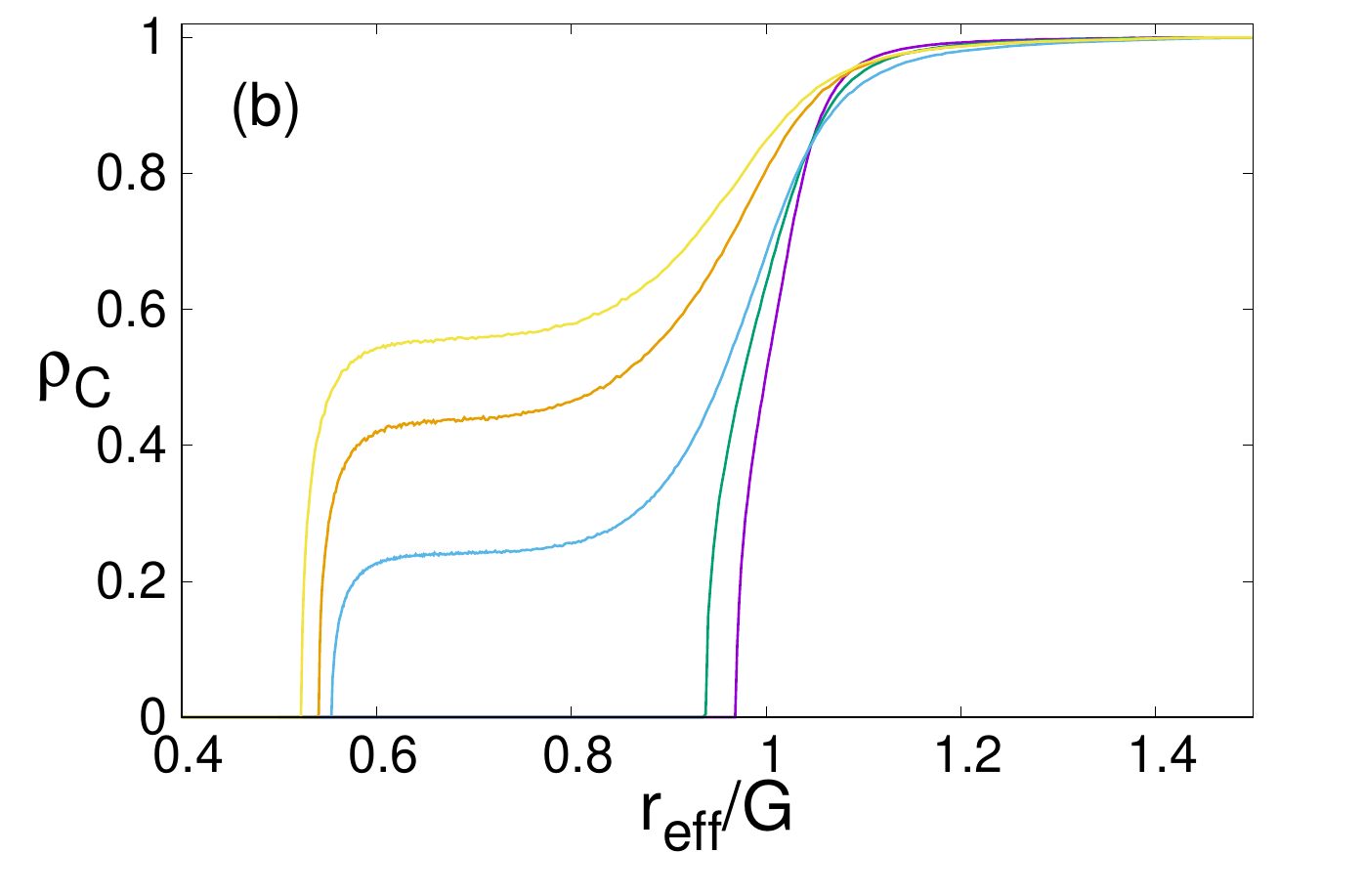}}
    \caption{(a) Asymptotic fraction of cooperators for a Prisoner's dilemma with $T=1+\gamma$, $R=1$, $P=0$, $S=-\gamma$, as function of $\gamma$.  (b) The PD can be imperfectly mapped onto the FPGG with a fixed contribution $c$, by using $r_{eff}=(1+\gamma)/\gamma$. In this case, the behavior of both games is qualitatively the same.
     }
    \label{pd}
\end{figure}

Next, we show the asymptotic cooperation fraction from the PD with self interaction simulated with a specific case of the parametrization above, $T = 1+\gamma$, $R = 1$, $P = 0$ and $S = - \gamma$, in a relatively noisy scenario with $K=0.1$ fixed in different lattices in  Fig.~\ref{pd}~(a). A mapping to the FPGG will correspond to a less noisy case which depends on each topology. 
Thus, we can only map the PD with self-interaction onto the FPGG for different noises in each lattice.
However, if we compare the outcomes on different topologies and still want to maintain $c$ fixed, an imperfect mapping is still possible by defining an effective multiplicative factor following eq.~\ref{eq_rmap},
\begin{equation}
    r_{eff} = \frac{1+\gamma}{\gamma}.
\end{equation}
Although Fig.~\ref{pd}~(b) does not correspond to a FPGG, the behavior of $\rho_C$ as a function of $r_{eff}/G$ for the PD  nevertheless preserves the same features found in the FPGG, as shown in Fig~\ref{norm}~(b),  and both games display qualitatively similar properties.

Finally,  we remark that the parameter space of the PD is of higher dimensionality than that of the FPGG; therefore it is expected that such mappings between the two games are not always possible. As an example, the typically used weak dilemma parametrization, $T = b$, $R = 1$, $P = S = 0$ with $b>1$ cannot be transformed into a FPGG, because it would yield a null contribution.

\section{Summary}

The question concerning the emergence and maintenance of cooperation has been extensively studied using many approaches and fields in the last years. Different mechanisms such as direct and indirect reciprocity, kin and group selections, and diversity, among others have been proposed to further understand how cooperation can survive in a competitive and egotistic environment. 
One of the most significant mechanisms in spatial reciprocity emerges spontaneously from the trivial fact that natural agents are generally confined to a finite region and only interact locally. 
For this reason, a better understanding of how diverse topologies can affect cooperation is of paramount importance to its understanding. 
In the current work, we studied three different dilemmas games: the classic Public Goods, the Focal Public Goods, and the Prisoner's Dilemma. We ran evolutionary simulations of all three using  Monte Carlo methods in the most usual two-dimensional regular topologies, i.e., the square, triangular, kagome and honeycomb lattices. For completeness, we also compared the results with the cubic and 4D hypercubic lattices.

First, we analyzed the random neighborhood case, where all group members were randomly selected among the population at each time step, suppressing spatial reciprocity. In this scenario, cooperators cannot coexist with defectors and only survive under the weak altruism condition, $r/G>1$, for the FPGG and the PGG.
On the other hand, coexistence is possible on the regular lattices, since now cooperators can avoid exploitation by forming compact groups. We found that topologies with larger group sizes, $G$,  are the most beneficial for cooperation in both the FPGG and PGG games.
Furthermore, cooperation can be enhanced by allowing shared neighbors between two connected players for the FPGG.

We propose a representation of the classic PGG as a focal game with an extended neighborhood where each site is weighted accordingly to its layer defined by the specific lattice topology. This allows us to obtain a single analytical framework to represent any regular lattice as different weighting schemes for the focal player's payoff. We used this scheme to point out important microscopic differences between the kagome lattice and the square lattice with von Neumann neighborhood for the PGG and show that a more clustered network does not necessarily favor cooperation. To our knowledge, this is the first time this approach has been presented.
In this game, the location of the $C$ neighbors in the cluster becomes relevant since closer neighbors participate in more than one game together. 
Therefore, for lattices with the same $G$ and a different number of shared connections (clustering coefficient), cooperation can be enhanced or inhibited depending on the situation. 

We also showed that a mapping between the PD and FPGG games is sometimes possible, depending on the parametrization chosen for the PD. When possible it is lattice dependent, meaning that a different value of the contribution in the FPGG is necessary for each lattice. However, if a fixed contribution is used, it is still possible to define an imperfect mapping with an effective multiplicative parameter in the FPGG.

Spatial reciprocity is a powerful enhancer of cooperation. In this work, we explored how diverse regular connection topologies can enhance cooperation in different games. But there are also known situations where spatial reciprocity inhibits cooperation. For example, the use of the aspiration rule for the updating of strategies~\cite{Du2015} found no improvement from the well-mixed to a spatially structured case. Other than that,  it was found that the introduction of spatial structure can even jeopardize cooperation in the snowdrift game~\cite{Hauert2004}.
The effect of group size was already studied~\cite{Wang2012a, Szolnoki2011,Szabo2009,Szolnoki2011} showing that an increase in the group size, for a fixed lattice, is beneficial for cooperation. 
This benefit ceases at a certain threshold where clustering to avoid exploitation is overcome by the length of interactions, reaching a well-mixed scenario.
As we can see, while spatial reciprocity is one of the strongest mechanisms to promote the spontaneous emergence of cooperation, many intricacies still need to be studied, especially in situations involving different topologies.

\section*{CRediT authorship contribution statement}
Lucas S. Flores: Conceptualization, Software, Formal analysis,
Writing - original draft. Heitor C.M. Fernandes: Conceptualization,
Formal analysis, Writing - original draft. Marco A. Amaral: Conceptualization, Formal analysis, Writing - original draft. Mendeli H. Vainstein: Conceptualization, Software, Formal analysis, Writing - original draft.

\section*{Declaration of Competing Interest}
The authors declare that they have no known competing financial interests or personal relationships that could have appeared to influence the work reported in this paper.

\section*{Acknowledgements}
L.S.F. thanks  the Brazilian funding agency CAPES (Coordenação de Aperfeiçoamento de Pessoal de Nível Superior) for the Ph.D. scholarship. 
M.H.V. acknowledges his ICTP associateship.
The simulations were performed on the IF-UFRGS computing cluster infrastructure.

\section*{Appendix} 
\label{app}

  \begin{figure*}[!ht]
     \centering
     \includegraphics[width=\linewidth]{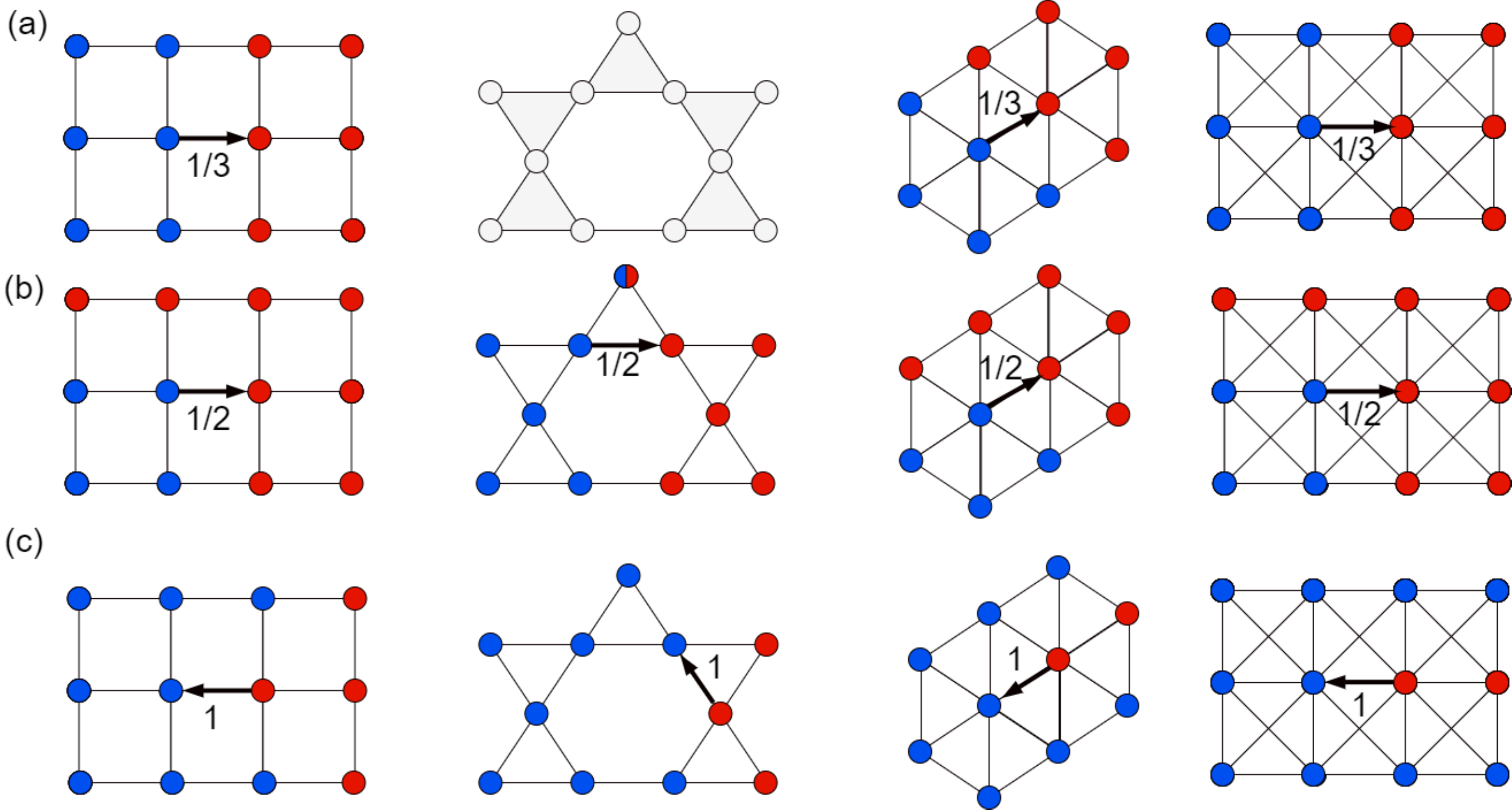}
     \caption{ Microscopic configurations of cooperators (blue) and defectors (red) that illustrate possible expansions or invasions of a cooperator cluster for regular lattices (square lattice with von Neumann neighborhood, kagome lattice, triangular lattice, and square lattice with Moore neighborhood). The arrows indicate possible cluster expansions (a)  along edges, with $r/G > 1/3$, (note that there is no transition with this value for the kagome lattice)  (b) and along vertices, with $r/G>1/2$. The top site with both colors in the kagome lattice indicates that this transition is independent of that site's strategy.  Although some expansions become possible at $r^*=1/3$, they alone are not sufficient for cooperation to survive. In (c), we display cluster invasions by defectors that are possible when  $r/G < 1$. Therefore, in the region $1/2< r/G < 1$,  the existence of both expansions and invasions with similar frequencies is what allows dynamical coexistence between the two strategies.
      }
     \label{retas}
\end{figure*}

Here we explore the conditions for the survival of a cooperator cluster in a sea of defectors for the FPGG in the deterministic regime, and also observed in low noise simulations shown in the inset of Fig.~\ref{norm}~(b).
We illustrate a few microscopic situations in Fig.~\ref{retas} for possible strategy flips on the square lattice with von Neumann and Moore neighborhoods, the kagome, and triangular lattices.
In a spatial lattice, cooperators survive by clustering; however, cooperators in contact with the sea of defectors will only survive if the inner $C$'s compensate their exploitation, i.e., $N^{{\scriptscriptstyle(C)}}_C \geqslant N^{{\scriptscriptstyle(D)}}_C$ (see eq.~\ref{rG-limit}). 
Figure~\ref{retas}~(a) shows the expansions due to cooperators at the borders of a cluster. Although these edge expansions become possible at and above $r^*/G=1/3$ (but below $r^*/G=1/2$), they occur at a much lower frequency than all possible invasions by defectors and, therefore, cooperation is not supported at low values of $r$. We notice that there is no such transition at this value for the kagome lattice. 
Cooperation can become viable when clusters survive both border and corner invasions and expand, which can only happen for $r/G>1/2$, as shown in Fig.~\ref{retas}~(b). 
If $r/G>1$, we are in a region where a cooperator with $N^{{\scriptscriptstyle(C)}}_C \geqslant N^{{\scriptscriptstyle(D)}}_C$  survives all possible invasions and expand, dominating the population for all topologies.
The interesting region is when $1/2 < r/G<1$, because as the cluster expands, some invasions are possible and cooperator dominance cannot occur. Examples of possible invasions for each lattice are shown in Fig.~\ref{retas}~(c).
The simultaneous existence of expansions and invasions is what drives the dynamical equilibrium in this parameter region, which allow clusters to change shape  while preserving the total cooperator density.

For the square lattice with the von Neumann neighborhood, the invasions outnumber the expansions during the system evolution and cooperation becomes extinct in a low noise scenario.  Despite sharing the same group size, $G=5$, the kagome lattice and the von Neumann neighborhood differ in that the former has common links between two connected players, as discussed in Sec.~\ref{sec.topologies}, thus being more clustered in this sense.
Therefore a cluster of $C$'s can expand in the kagome lattice, with support from the inner $C$'s, what   
 explains the coexistence region in Fig.~\ref{norm}.

As we change lattices, increasing group size, there will be more configurations that generate the same transition values $r^*/G$, because the same values of $\Delta N_C $ can be achieved with more sets of $N^{{\scriptscriptstyle(C)}}_C$ and $N^{{\scriptscriptstyle(D)}}_C$. Such configurations can be favorable or detrimental to cooperation. However, only those that satisfy $N^{{\scriptscriptstyle(C)}}_C \geqslant N^{{\scriptscriptstyle(D)}}_C$, the ones related in general to clustered cooperators,  will matter as the system evolves. As a result, cooperators survive more configurations in the more connected lattices.
Thus, despite there being similar invasion in the von Neumann neighborhood, triangular lattice and Moore neighborhoods, cooperator clusters will be able to expand and survive due to the increase in connections in the latter two. 

An analogous analysis of microscopic configurations can also be made for the PGG, considering that it can be mapped to a FPGG with an extended neighborhood. However, since the group sizes and number of microscopic configurations become much larger, we believe that displaying them would be more confusing than helpful.


\end{document}